\documentclass[useAMS,usenatbib]{mn2e}
\usepackage{graphicx}
\usepackage{epstopdf}
\usepackage[figuresright]{rotating}
\usepackage{subfigure,color}
\usepackage{longtable}
\usepackage{xspace}

\newcommand{\ion}[2]{{\textrm{#1}}\,{\textrm{\sc #2}}}

\definecolor{pink}{rgb}{.9,.2,.5}  
\definecolor{purple}{rgb}{.5,.6,.7}

\title[Abundances in the central region of galaxies]
{On the central abundances   of Active Galactic Nuclei and Star-forming  Galaxies}

\author[Dors et al.]
            { 
              O.~L.\ Dors Jr.$^{1}$\thanks{E-mail:olidors@univap.br}, 
              M.~V. Cardaci$^{2,3}$, 
              G.~F.\ H\"agele$^{2,3}$,  
              I. Rodrigues$^{1}$, 
 \newauthor{  E.~K.\ Grebel$^{4}$, 
              L.~S.\ Pilyugin$^{4,5,6}$, 
              P.\ Freitas-Lemes$^{1}$, 
              A.~C.\ Krabbe$^{1}$, 
             } \\
$^1$ Universidade do Vale do Para\'iba, Av. Shishima Hifumi, 2911, Cep
12244-000, S\~ao Jos\'e dos Campos, SP, Brazil\\ 
$^2$ Instituto de Astrof\'isica de La Plata (CONICET-UNLP), Argentina. \\
$^3$ Facultad de Ciencias Astron\'omicas y Geof\'{\i}sicas, Universidad Nacional de La Plata, Paseo del Bosque s/n, 1900 La Plata, Argentina.\\
$^4$ Astronomisches Rechen-Institut, Zentrum f\"ur Astronomie der Universit\"at Heidelberg, M\"onchhofstr.\ 12-14, 69120 Heidelberg, \\
Germany\\
$^5$ Main Astronomical Observatory of National Academy of Sciences of Ukraine, 27 Zabolotnogo str., 03680 Kiev, Ukraine\\
$^{6}$ Kazan Federal University, 18 Kremlyovskaya St., 420008, Kazan, Russian Federation \\
}

\begin{document}

\date{Accepted 2015 Month  00. Received 2015 Month 00; in original form 2014 December 17}

\pagerange{\pageref{firstpage}--\pageref{lastpage}} \pubyear{2011}

\maketitle

\label{firstpage}

\begin{abstract}
We examine the relation between oxygen abundances in the narrow-line regions (NLRs) of active galactic nuclei (AGNs) 
estimated from the optical emission lines through the strong-line method (the theoretical  calibration of \citealt{thaisa98}), 
via the direct $T_e$-method, and the central intersect abundances in the host galaxies determined from the radial abundance gradients.
We found that the $T_e$-method underestimates the oxygen abundances by up to $\sim$2 dex 
(with average value of $\sim 0.8$ dex) compared to the abundances
derived through the strong-line method. This confirms the existence of the so-called ``temperature problem'' in AGNs. 
We also found that the abundances in the centres of galaxies obtained from their spectra trough the strong-line method
are close to or slightly lower than the central intersect abundances estimated from the radial abundance gradient 
both in  AGNs and   Star-forming galaxies.
The oxygen abundance of the NLR is usually lower than the maximum attainable abundance in galaxies ($\sim$2 times the solar value).  
This suggests that there is no extraordinary chemical enrichment of the NLRs of AGNs.
\end{abstract}

\begin{keywords}
galaxies: general -- galaxies: evolution -- galaxies: abundances --
galaxies: formation-- galaxies: ISM
\end{keywords}


\section{Introduction}

 Active galactic nuclei (AGNs) and star-forming regions show 
prominent  emission lines  of heavy elements that can be easily measured, even for objects at large redshifts. 
The intensity of these emission lines  depends on  the metallicity, 
 which makes them convenient tracers of
the chemical evolution of the Universe.

The abundance of a given element in AGNs and in star-forming regions can be derived from measurements of the
relative strengths of the emission lines of its ions and of the  electron temperature  and density of the gas 
\citep{osterbrock06}.  Oxygen has been  generally used as tracer  of the metallicity ($Z$) of star-forming regions 
\citep[e.g.,][]{tremonti04,Thuan2010, nagao06a} and of AGNs \citep[][among others]{izotov10, groves06a}.
Methods for the abundance determinations in star-forming regions have been discussed 
in many papers \cite[e.g.,][]{stasinska04}.  The consensus is  that {\it bona-fide} determinations 
of $Z$  are only  obtained   by using the $T_{\rm e}$-method \citep{hagele06,hagele08}.   This method  
is based on the determination of the electron temperature ($T_{\rm e}$) from the ratios between intensities of two  
emission lines originating in transitions from two levels 
with different excitation energies of a same ion, such as the ratios [\ion{O}{iii}]($\lambda$4959+$\lambda$5007)/$\lambda$4363,
[\ion{S}{iii}]($\lambda$9069+$\lambda$9532)/$\lambda$6312, and  [\ion{N}{ii}]($\lambda$6548+$\lambda$6584)/$\lambda$5755
(see \citealt{osterbrock06}.)
 Unfortunately, these  line ratios cannot easily be measured  in distant objects and/or in  objects with low excitation
(e.g., \citealt{bresolin05}),  preventing the  use of the $T_{\rm e}$-method.

 Therefore, the oxygen abundances in H\,{\sc ii} regions are 
usually estimated using the strong-line method pioneered by \citet{pagel79} and
\citet{alloin79}.  The principal idea of the strong-line method is
to establish the relation between the (oxygen) abundance in an H\,{\sc ii} region and some combination of the 
intensities of strong emission lines in its spectrum, i.e., the position of the objects in the strong-line diagram 
is calibrated in terms of abundance. Therefore such a relation is often called a  ``calibration''. 
 The calibrations of strong emission lines in terms of the oxygen abundances  can be accomplished by three methods: 
(i) theoretical calibrations based on  photoionization models \citep[e.g.,][]{kewley02,dors11, dors05}, 
(ii) empirical calibrations based on H\,{\sc ii} regions with  abundances derived through the $T_{\rm e}$-method 
\citep[][among others]{leonid00,pilyugin01,pilyugin2005,perez09}, and 
(iii) hybrid calibrations where both   photoionization models and  H\,{\sc ii} regions with $T_{e}$-based 
abundances are used as calibration data points \citep{pettini04}. The metallicity scales produced 
by these different methods may differ from each other  by up 0.7 dex  \citep{kewley08}. 

 Here we use the metallicity scale in 
H{\sc ii} regions defined by the abundances determined through the classical $T_{e}$-method.
This is because in the $T_{e}$ method
the physical conditions in the nebulae, which are essential ingredients in order to calculate the abundance, are derived 
 directly from observations. 
 Otherwise, if the physical conditions and consequently the abundances are derived by models they will be less certain.
 This is because of the parameters of the model such as the ionizing source,  geometry, etc., 
tend not to be sufficiently realistic (see \citealt{dors05, kennicutt03, viegas02}).

AGN metallicities are usually estimated through strong-line theoretical calibrations  \citep[e.g.,][]{thaisa98}. 
Even solar or supersolar abundances have been  found. \citet{richardson14} established a sequence of photoionization models to 
reproduce the   narrow-line regions (NLRs) of AGN spectra and found  that models with a  metallicity  of  $1.4 \: Z_{\odot}$ 
provide the  best  agreement  with  the   observational data. \citet{groves06a} used the photoionization models to analyse 
the emission lines in the spectra of the NLRs from the Sloan Digital 
Sky Survey (see \citealt{tremonti04}) and found supersolar metallicities  for typical Seyfert galaxies (see also 
\citealt{batra14, du14, wang11, dhanda07, baldwin03, hamann02, thaisa98, ferland96, hamann93, hamann92}). 
High metallicities at the centres of spiral galaxies  are also obtained  by the  extrapolation of  radial abundance 
gradients to the  central  regions \citep[][among others]{VilaCostas1992,Zaritsky1994,vanZee1998,pily04,pilyugin2007,Gusev2012}.
 
In contrast,  low metallicities in AGNs have been obtained when the  $T_{\rm e}$ method was used \citep[e.g.,][]{zhang13}.  
The disagreement between the abundances obtained through the strong-line methods and through the $T_{\rm e}$-method 
is the  so-called ``temperature problem''. 
On the other hand, \citet{izotov08}  determined the abundances through the  $T_{\rm e}$-method in the AGNs in four dwarf galaxies
and found metallicities between 0.05 and 0.2 of the solar value, which is a typical abundance of those galaxies. 
\citet{alloin92},  also using the $T_{\rm e}$-method,
estimated the  abundance in the Seyfert 2 nucleus of the  galaxy ESO\,138\,G1 classified as E/S0 \citep{lauberts82} and found $Z\approx0.4 \: Z_{\odot}$.
Thus, the  $T_{\rm e}$ method seems to produce realistic abundance estimations in the low metallicity range, i.e., 
the temperature problem seems to disappear at low metallicities. 
Finding the correct method for the determination of AGN metallicities and 
establishing a relation between the AGN metallicity and the metallicity of the host galaxy are very important challenges.   
It should be noted that the AGN metallicity is sometimes adopted as a
surrogate metallicity  of the host galaxy, mainly at high redshifts (e.g., \citealt{dors14, matsuoka09, nagao06b}).  
 
Nowadays, a large number  of  measured intensities of emission lines sensitive to the electron temperature are available for
AGNs in the literature.  This also holds for determinations of oxygen gradients in a large sample of spiral galaxies.
This provides the possibility for an analysis of the metallicity determinations in AGNs.
 The main goals of the current study  are:
 (i) To investigate the discrepancy between AGN abundances derived through the strong-line methods and through the  $T_{\rm e}$ method. 
The magnitude of this discrepancy is well known for star-forming regions \citep[][among others]{pilyugin03,kewley08,lopez10}, 
 however, this problem has received less attention in the case of the NLRs of AGNs. 
(ii)  To compare the abundances of the  NLRs of AGNs and nuclear star-forming regions 
estimated from their spectra and central oxygen abundances of the host galaxies 
derived from the radial abundance gradients. Differences between these 
values could be evidence in favour of the existence of gas infall onto the center of a galaxy or extraordinary chemical 
enrichment of  these objects.

The present study  is organised as follows. In  Section 2 we compare the NLR abundances 
determined through the $T_{e}$-method and the strong emission-line relations of \citet[][hereafter SB98]{thaisa98}.   
We examine the relation between the abundances in the NLRs estimated from their optical emission-line spectra 
and the central intersect abundances in the host galaxies determined from the radial abundance gradients in Section 3. 
The conclusions are given in Section 4.

\section{Abundance determinations in NLRs: $T_{e}$-method vs. relations by  SB98}

\subsection{AGN sample}
\label{sdata}

Intensities of  narrow emission lines of AGNs were compiled from the literature. Our selection criterion was 
the presence of flux measurements of the narrow optical emission lines [\ion{O}{ii}]$\lambda$3727, [\ion{O}{iii}]$\lambda$4363, 
H$\beta$, [\ion{O}{iii}]$\lambda$5007, H$\alpha$, [\ion{N}{ii}]$\lambda$6584, and
 [\ion{S}{ii}]$\lambda\lambda$6717,6731. 
We only considered galaxies whose  nuclei were classified  as  Seyfert 2 (Sy2) and Sy1.9. 
 Sequences of photoionization models, not including the  shock gas but only 
a power law radiation of the ionizing source,  are able to reproduce measurements of strong  
narrow emission lines of Sy2 and Sy1.9 for a large spectral range (e.g., \citealt{riffel13a, dors12, groves06a}). 
Hence, the  selection of these  AGN types  minimises the effects of shock gas excitations and ionizations, 
which are not considered in our abundance determinations.
 
We list in   Table~\ref{tab1} the identification of the objects,  the emission-line intensities 
(relative to H$\beta$=1.0), the spectral classification taken from the 
{\sc ned}\footnote{The NASA/IPAC Extragalactic Database ({\sc ned}) is operated by the Jet 
Propulsion Laboratory, California Institute of Technology, under contract with the National Aeronautics and 
Space Administration.},  and bi\-blio\-graphic references from which  the data were taken. The sample consists 
of  47 local AGNs (redshift $z<0.1$) observed with long-slit spectroscopy. 
In the cases where the reddening correction was not performed in the original works (indicated in Table.~\ref{tab1}), 
we dereddened the emission-lines comparing the observed  H$\alpha$/H$\beta$ ratio with the theoretical value of 2.86  
\citep{hummer87},  obtained for an electron temperature of 10\,000 K and an electron density of $100\: \rm cm^{-3}$.  
  
The emission-line flux errors are   reported in the original works for only 4 out of the 47 objects of our sample. 
Therefore the errors are not given in Table~\ref{tab1}.
Typical errors  of the emission-line intensities are about 10 to 20 per cent (e.g., \citealt{kraemer94}), which 
yield  uncertainties in the oxygen abundance estimations   of about 0.1 to 0.2 dex (e.g., \citealt{rosa14, hagele08, dors11, kennicutt03}). 
 Hereafter we will assume that the oxygen estimations from the $T_{e}$-method and the relations by
SB98 have an uncertainty of 0.15 dex, an average of the values above. 
 For the objects  which the flux of the  line   [\ion{O}{iii}]$\lambda$4959 is not available,
this was calculated from the theoretical relation $I$[\ion{O}{iii}]$\lambda$4959=$I$[\ion{O}{iii}]$\lambda$5007/3.0.

\begin{table*}
\caption{Emission-line intensities of AGNs relative to H$\beta$=1.00 compiled from the literature.}
\label{tab1}
\begin{tabular}{|lcccccccccc}	 
\noalign{\smallskip} 
\hline 
 Object           &  [\ion{O}{ii}]$\lambda$3727  &  [\ion{O}{iii}]$\lambda$4363 &[\ion{O}{iii}]$\lambda$4959 & [\ion{O}{iii}]$\lambda$5007   & H$\alpha$ & [\ion{N}{ii}]$\lambda$6584 & [\ion{S}{ii}]$\lambda$6717 & [\ion{S}{ii}]$\lambda$6731 & Type     & Ref.\\
\hline
NGC\,7674                &           1.29               &           0.12                        &                3.94                   &         12.55                 &    3.70   &         3.68               &         0.54               &            0.64            & Sy2      &  1 \\
IZw\,92                      &           2.63               &           0.32                        &                3.52                   &         10.12                &    2.42   &         0.97               &         0.37               &            0.40            & Sy2      &  1 \\                                
NGC\,3393$^{(a), (b)}$ &           2.41               &           0.14                    &                 5.47                    &         16.42                 &    2.78   &         4.50               &         0.64               &            0.89            & Sy2      &  2 \\
Mrk\,176                   &           3.54               &           0.32                        &                4.69                     &         14.36                 &    2.81   &         2.99               &         0.56               &            0.54            & Sy2      &  3 \\
3c33                         &           4.93               &           0.32                        &                4.10                    &         12.68                 &    2.63   &         1.76               &         0.87               &            0.73            & Sy2      &  3 \\
Mrk\,3                      &           3.52               &           0.24                         &                3.99                    &         12.67                  &    3.10   &         3.18               &         0.73               &            0.82            & Sy2      &  3 \\  
NGC\,1068               &           1.23               &           0.22                         &                4.11                    &         12.42                  &    2.57   &         4.55               &         0.26               &            0.55            & Sy2      &  3 \\
Mrk\,573                  &           2.92               &           0.18                         &                 3.89                    &         12.12                  &    2.95   &         2.47               &         0.75               &            0.80            & Sy2      &  3 \\
Mrk\,78                   &           4.96               &           0.14                          &                 3.88                   &         11.94                   &    2.46   &         2.32               &         0.68               &            0.61            & Sy2      &  3 \\
Mrk\,34                   &           3.43               &           0.15                          &                 3.68                   &         11.46                   &    2.99   &         2.18               &         0.82               &            0.80            & Sy2      &  3 \\
Mrk\,1                     &           2.78               &           0.21                          &                3.56                     &         10.95                   &    2.66   &         2.21               &         0.49               &            0.52            & Sy2      &  3 \\
3c433                     &           6.17               &           0.41                           &                3.27                     &          9.44                    &    3.38   &         5.13               &         1.58               &            1.13            & Sy2      &  3 \\              
Mrk\,270                &           5.64               &           0.28                            &                2.96                     &          8.71                     &    3.14   &         2.93               &         1.21               &            1.39            & Sy2      &  3 \\
3c452                   &           4.81               &           0.18                            &                 2.40                    &          6.85                      &    2.98   &         3.58               &         1.10               &            0.77            & Sy2      &  3 \\              
Mrk\,198               &           2.51               &           0.12                            &                 1.79                    &          5.56                      &    3.02   &         2.26               &         0.89               &            0.68            & Sy2      &  3 \\            
Mrk\,268               &           3.75               &           0.25                            &                 1.55                    &          4.82                      &    3.38   &         4.94               &         1.28               &            1.08            & Sy2      &  3 \\   
Mrk\,273               &           8.27               &           0.22                            &                 1.52                    &          4.44                      &    2.94   &         2.62               &         0.89               &            0.54            & Sy2      &  3 \\                 
NGC\,3227$^{(b)}$ &           3.22             &           0.50                            &                  3.57                    &         10.73                     &    2.86   &         5.01               &         1.24               &            1.26            & Sy2      &  4 \\                
Mrk\,6                   &           2.45               &           0.28                           &                  3.37                   &         10.13                    &    2.79   &         1.79               &         0.62               &            0.63            & Sy2      &  4 \\                    
ESO\,138\,G1       &           2.35               &           0.34                            &                  2.93                   &          8.71                      &    3.01   &         0.68               &         0.47               &            0.48            & Sy2      &  5 \\
NGC\,5643$^{(a),(b) }$ &   5.55               &           0.54                            &                  4.20                    &         12.61                     &    2.66   &         2.90               &         0.91               &            0.71            & Sy2      &  6 \\
NGC\,1667            &        11.50               &           0.59                            &                  2.78                    &          9.20                      &    2.80   &         6.96               &         1.37               &            1.17            & Sy2      &  7 \\
Mrk\,423               &           8.00               &           0.40                            &                 2.80                    &          6.20                      &    2.90   &         3.60               &         1.20               &            1.00            & Sy1.9    &  8 \\                        
Mrk\,609               &           1.80               &           0.50                            &                 1.80                     &          5.00                      &    2.80   &         2.60               &         0.60               &            0.50            & Sy1.9    &  8 \\
Mrk\,226SW         &           5.23               &           0.08                             &                 1.50                    &          4.50                     &    3.35   &         2.07               &         0.29               &            0.25            & Sy2      & 9 \\
NGC\,3081$^{(a)}$ &         2.16               &           0.23                             &                 4.53                    &         12.62                     &    2.73   &         2.33               &         0.60               &            0.62            & Sy2      & 10 \\ 
NGC\,3281$^{(a)}$  &        2.33               &           0.22                             &                 2.77                     &          7.59                     &    2.64   &         2.54               &         0.53               &            0.60            & Sy2      & 10 \\
NGC\,3982$^{(a)}$   &      3.92               &           0.03                              &                 5.64                     &         18.68                     &    2.76   &         2.57               &         0.76               &            0.82            & Sy2      & 10 \\
NGC\,4388$^{(a)}$   &      2.68               &           0.15                              &                 3.67                      &         10.63                     &    2.71   &         1.44               &         0.68               &            0.60            & Sy2      & 10 \\
NGC\,5135$^{(a)}$   &      2.01               &           0.10                              &                 1.40                      &          4.47                      &    2.64   &         2.35               &         0.37               &            0.35            & Sy2      & 10 \\
NGC\,5643$^{(a)}$   &     5.11               &           0.42                               &                  4.56                      &         15.4                        &    2.64   &         3.07               &         0.97               &            0.90            & Sy2      & 10 \\
NGC\,5728$^{(a)}$   &     3.41               &           0.44                               &                  3.70                      &         10.98                      &    2.65   &         3.71               &         0.41               &            0.41            & Sy2      & 10 \\
NGC\,6300$^{(a)}$   &   15.48               &           1.39                               &                  8.32                      &         23.32                      &    2.59   &         6.62               &         1.42               &            1.26            & Sy2      & 10 \\
NGC\,6890$^{(a)}$   &    2.77               &           0.72                                &                  7.48                      &         20.05                       &    2.75   &         4.26               &         0.65               &            0.52            & Sy2      & 10 \\
IC\,5063$^{(a)}$    &       5.06               &           0.28                                &                  3.37                      &         10.31                       &    2.77   &         2.67               &         0.69               &            0.60            & Sy2      & 10 \\
IC\,5135$^{(a)}$    &       4.05               &           0.25                                &                  2.07                       &          6.88                        &    2.65   &         3.30               &         0.49               &            0.46            & Sy2      & 10 \\
Mrk\,744           &           2.38               &           0.33                                &                   3.18                       &          8.84                        &    2.47   &         3.62               &         2.83               &            2.83            & Sy2      & 11 \\                       
Mrk\,1066          &          3.34               &           0.08                                &                   1.22                      &          3.84                        &    2.76   &         2.42               &         0.51               &            0.55            & Sy2      & 11 \\
NGC\,5506          &        2.84               &           0.14                                &                   2.46                      &           7.69                        &    2.84   &         2.53               &         0.92               &            0.99            & Sy1.9    & 12 \\                                                                                           
NGC\,2110          &        4.38               &           0.26                                &                   1.61                      &          4.76                        &    2.66   &         3.76               &         1.52               &            1.38            & Sy2      & 12 \\
NGC\,3281$^{(a), (b)}$ &   4.30          &           0.42                                 &                   2.44                      &          7.34                        &    2.60   &         2.60               &         0.78               &            0.73            & Sy2      & 13 \\
Akn\,347$^{(a)}$    &      2.98               &           0.42                                &                   4.95                      &         15.01                       &    2.65   &         3.23               &         0.75               &            0.75            & Sy2      & 14 \\
UM\,16$^{(a)}$      &      2.90               &           0.22                                &                   4.62                      &         14.00                       &    2.72   &         1.70               &         0.45               &            0.45            & Sy2      & 14 \\
Mrk\,533$^{(a)}$    &     1.59               &           0.13                                 &                   4.03                      &         12.23                        &    2.72   &         2.72               &         0.39               &            0.45            & Sy2      & 14 \\
IZw\,92$^{(a)}$     &      2.60               &           0.34                                 &                   3.54                      &         10.14                        &    2.78   &         1.00               &         0.40               &            0.43            & Sy2      & 14 \\  
Mrk\,612$^{(a)}$    &     1.88               &           0.17                                 &                   2.99                      &          9.37                         &    2.67   &         3.60               &         0.74               &            0.55            & Sy2      & 14 \\
Mrk\,622$^{(a)}$    &    10.06              &           0.03                                 &                   1.90                       &          5.44                          &    2.47   &         2.33               &         0.19               &            0.14            & Sy2      & 14 \\
\hline 
\end{tabular}
\begin{minipage}[c]{2\columnwidth}
References--- (1) \citet{kraemer94},  (2) \citet{contini12},  (3) \citet{koski78},  (4) \citet{cohen83}, (5) \citet{alloin92} \\
(6)  \citet{enrique94} (7) \citet{radovich96}, (8) \citet{osterbrock81}, (9) \citet{osterbrock83}, \\ (10) \citet{phillips83}, (11) \citet{goodrich83},
 (12) \citet{shuder80},  (13) \citet{durret88}, \\ and (14) \citet{shuder81}.  
$^{(a)}$ Data corrected for reddening in this work. $^{(b)}$  Value of  $I$[\ion{O}{iii}]$\lambda$4959
estimated from the theoretical relation $I$[\ion{O}{iii}]$\lambda$4959=$I$[\ion{O}{iii}]$\lambda$5007/3.0.\\
\end{minipage}
\end{table*}

\subsection{Relations of SB98} 
\label{thaisasec}

SB98 carried out NLR model calculations using the photoionization code {\sc Cloudy} \citep{ferland96} and 
suggested two relations for the abundance determinations in the NLRs of AGNs. The first one is 
\begin{eqnarray}
       \begin{array}{lll}
{\rm (O/H)}_{{\rm SB98,1}}& = &  8.34  + (0.212 \, x) - (0.012 \,  x^{2}) - (0.02 \,  y)  \\  
         & + & (0.007 \, xy) - (0.002  \, x^{2}y) +(6.52 \times 10^{-4} \, y^{2}) \\  
         & + & (2.27 \times 10^{-4} \, xy^{2}) + (8.87 \times 10^{-5} \, x^{2}y^{2}),   \\
     \end{array}
\label{equation:sb1}
\end{eqnarray}
where $x$ = [N\,{\sc ii}]$\lambda$$\lambda$6548,6584/H$\alpha$ and 
$y$ = [O\,{\sc iii}]$\lambda$$\lambda$4959,5007/H$\beta$.  
The second relation of SB98 is 
\begin{eqnarray}
       \begin{array}{lll}
{\rm (O/H)}_{{\rm SB98,2}}   & = &  8.643 - 0.275 \, u + 0.164 \, u^{2}   \\  
          & + & 0.655 \, v - 0.154 \, u v  - 0.021 \, u^{2}v \\  
          & + & 0.288 v^{2} + 0.162 u v^{2} + 0.0353 u^{2}v^{2},   \\
     \end{array}
\label{equation:sb2}
\end{eqnarray}
where $u$ = log([O\,{\sc ii}]$\lambda$$\lambda$3727,3729/[O\,{\sc iii}]$\lambda$$\lambda$4959,5007) 
and $v$ = log([N\,{\sc ii}]$\lambda$$\lambda$6548,6584/H$\alpha$).  Both calibrations
are valid for   $\rm 8.4 \: \lid \: 12+log(O/H) \:  \lid \: 9.4$.

The dependence of these relations on the density should be taken into account. This dependence is given
by the expression  considered by SB98
\begin{equation}
\label{necorr}
({\rm O/H})_{{\rm final}} = {\rm O/H} - 0.1\,\,\log(N_{e}/300), 
\end{equation}
where $N_{e}$ is the electron density in cm$^{-3}$ and the correction is valid for 100 cm$^{-3}$ $\la$ $N_{e}$ 
$\la$ 10$^{4}$ cm$^{-3}$. It should be noted that the value of this correction exceeds 0.1 dex for 
high-density objects only, i.e., for objects with electron densities $N_{e}$ $\ga$ 3$\times$10$^{3}$ cm$^{-3}$.

\subsection{$T_{\rm e}$-method}
\label{temed}

We use the emission-line intensities listed in Table~\ref{tab1} to determine the oxygen abundances (O/H)$_{T_{e}}$ 
of the narrow-line regions through the classic $T_{\rm e}$-method. 
We will follow the  methodology described in \citet{dors13}   based on the equations given 
by \citet{perez09}, \citet{hagele08}, \citet{perez07}, and \citet{perez03}. 

The electron temperature in the high ionization zone (referred to as $t_{3}$) for each object was obtained from  
the observed line-intensity ratio $R_{\rm O3}$=[\ion{O}{iii}]($\lambda4959\: + \: \lambda 5007)/\lambda4363$
using the expression 
\begin{equation}
 \label{eqt3}
 t_{3}=0.8254-0.0002415 R_{\rm O3}+\frac{47.77}{R_{\rm O3}},
 \end{equation}
with $t_{3}$ in units of $10^{4}$K. The electron temperature of the low ionization zone (referred to as $t_{2}$) was derived from
the theoretical relation: 
\begin{eqnarray}
\label{eqo3}
t_{2}^{-1}\,=\,0.693\,t_{3}^{-1}+0.281.
\end{eqnarray}
The $\rm O^{++}$  and $\rm O^{+}$ ionic abundances were computed  through the relations: 
\begin{eqnarray}
 12+\log(\frac{{\rm O^{++}}}{{\rm H^{+}}}) \!\!\!&=&\!\!\! \log \big( \frac{I(4959)+I(5007)}{I{\rm (H\beta)}}\big)+6.144  \nonumber\\
                                          &&\!\!\!+\frac{1.251}{t_{3}}-0.55\log  t_{3}.   
   \end{eqnarray}
 and
\begin{eqnarray}
 12+\log(\frac{{\rm O^{+}}}{{\rm H^{+}}})  \!\!\!&=&\!\!\! \log  \big( \frac{I(3727)}{I{\rm (H\beta)}}\big)+5.992 \nonumber\\
                                           &&\!\!\!+\frac{1.583}{t_{2}}-0.681\log t_{2} +\log(1+2.3 n_{\rm e}).
\end{eqnarray}

Finally, we assumed
\begin{eqnarray}
\rm
\frac{O}{H}=\frac{O^{+}}{H^{+}}+\frac{O^{++}}{H^{+}} .
\end{eqnarray}
for the determination of the total abundance. 
 
\subsection{Abundance results}
\label{res}

\begin{table*}
\caption{ Object name, electron density, electron  temperature, and oxygen abundance obtained via the $T_{\rm e}$-method 
and via the calibrations of \citet{thaisa98} (Equations~\ref{equation:sb1} and \ref{equation:sb2})
for the objects in our sample. Oxygen abundance value estimations outside the  validity interval of the
calibrations were not considered.}
\label{tab2}
\begin{tabular}{@{}lccccc@{}} 
\noalign{\smallskip}
\hline    
                              &                                                     &                         & \multicolumn{3}{c}{12+log(O/H)} \\
\cline{4-6} 		   
 Object                   &          $N_{\rm e}$ (cm$^{-3}$)       &      $T_{\rm e}$ ($10^{4}$K)  & $T_{\rm e}$-method  &     SB98,1 &  SB98,2   \\ 
 \noalign{\smallskip} 
\hline
NGC\,7674            &                            1148.0                &       1.14            &     8.49                     &     8.72            &  9.19         \\            
IZw\,92                 &                            822.0                   &       1.95            &     7.90                      &     8.44            &  8.67          \\ 
NGC\,3393           &                           2022.0                  &       1.09            &     8.69                      &     ---               & 9.24   \\              
Mrk\,176               &                          535.0                    &       1.61            &     8.21                      &     8.84            &  9.02  \\              
3c33                     &                         252.0                     &       1.72            &     8.14                      &     8.64            & 8.80   \\              
Mrk\,3                   &                         948.0                    &       1.49            &     8.26                      &    8.75             & 8.96   \\ 
NGC\,1068           &                         26993.0                 &       1.44            &     8.40                      &    ---             &  ---  \\ 
Mrk\,573              &                         781.0                     &       1.34            &     8.33                      &    8.65             &  8.92  \\ 
Mrk\,78                &                       370.0                       &       1.22            &     8.49                      &   8.74              & 8.88   \\ 
Mrk\,34                &                    546.0                          &       1.27            &     8.39                      &   8.61              & 8.85   \\ 
Mrk\,1                  &                    767.0                          &       1.50            &     8.17                       &  8.63               & 8.90 \\ 
3c433                  &                   50.0                             &       2.38            &     7.84                      &  8.96               & 9.04   \\ 
Mrk\,270              &                    1027.0                        &       1.97            &     7.99                      &   8.62              &  8.75  \\ 
3c452                  &                    50.0                           &       1.76            &     7.95                       &  8.80               & 8.95  \\ 
Mrk\,198              &                    111.0                         &       1.59            &     7.88                       &  8.61               & 8.84   \\ 
Mrk\,268              &                     260.0                        &       2.68            &     7.52                      &   8.76              &  8.93  \\ 
Mrk\,273              &                      50.0                         &       2.60            &     7.73                      &   8.67              &  8.72  \\ 
NGC\,3227          &                     647.0                         &       2.49            &     7.75                       &  8.98              & 9.15  \\  
Mrk\,6                 &                    647.0                          &       1.81            &     7.95                      &   8.55               & 8.84   \\ 
ESO\,138\,G1      &                    685.0                         &       2.22            &     7.73                      &   8.40               & 8.53  \\  
NGC\,5643          &                     141.0                        &       2.36            &     7.90                      &   8.86              & 8.96   \\ 
NGC\,1667          &                    281.0                         &       3.12            &     7.84                     &   9.15              &  9.06  \\ 
Mrk\,423              &                  239.0                           &       3.14            &     7.67                     &    8.73              &  8.80 \\ 
Mrk\,609              &                  239.0                          &       4.41            &     7.11                     &    8.62             &  8.92 \\ 
Mrk\,226SW        &                 296.0                           &       1.45            &     8.08                      &    8.52              &  8.60  \\  
NGC\,3081          &              693.0                              &       1.84            &     7.85                     &   8.60              &  8.91   \\  
NGC\,3281          &               974.0                             &       1.86            &     7.84                      &   8.61               &  8.90  \\               
NGC\,3982         &               819.0                              &       0.68            &     9.54                      &   8.87              &  8.99  \\           
NGC\,4388         &               343                                 &       1.31            &     8.30                      &  8.54               &  8.80  \\           
NGC\,5135         &                492.0                             &       1.61            &     7.78                     &  8.57               &  8.83   \\           
NGC\,5643         &                451.0                             &       1.79            &     8.17                     &   8.93              &  8.99  \\           
NGC\,5728        &                 606.0                             &       2.26            &     7.84                     &    8.86              &  9.05   \\           
NGC\,6300        &                 360.0                             &       2.96            &     8.11                     &   ---              & 9.17   \\           
NGC\,6890       &                   176.0                            &       2.11            &     8.06                     &   ---              &   9.35  \\           
IC\,5063           &                  311.0                             &       1.79            &     8.06                     &    8.51             &  8.65 \\       
IC\,5135           &                471.0                               &       2.12            &     7.78                     &   8.72              & 8.90  \\       
Mrk\,744           &                606.0                               &       2.16            &     7.76                     &   8.83              &  9.10  \\    
Mrk\,1066         &                839.0                               &       1.56            &     7.90                     &   8.54              &  8.69  \\     
NGC\,5506       &               809.0                                &       1.46            &     8.09                     &    8.60             &  8.85   \\      
NGC\,2110       &              395.0                                 &       2.78            &     7.54                     &   8.73              &  8.87 \\      
NGC\,3281       &               471.0                                &       2.87            &     7.61                     &  8.66               & 8.83   \\           
Akn\,347           &              606.0                                 &       1.82            &     8.09                     &  8.93               &  9.11   \\      
UM\,16             &             606.0                                  &       1.37            &     8.35                     &  8.59               &  8.87   \\      
Mrk\,533           &             1046.0                                &       1.18            &     8.44                     &   8.72              &   9.12   \\       
IZw\,92             &              805.0                                 &       2.02            &     7.87                     &  8.42               &  8.65   \\      
Mrk\,612           &             75.0                                    &       1.46            &     8.10                    &   8.89              &  9.24  \\       
Mrk\,622           &             64.0                                    &       0.97            &     8.76                    &   8.69              &   8.73   \\       
\hline
\end{tabular}
\end{table*}

 Electron  densities ($N_{\rm e}$), \ion{O}{iii} electron temperatures ($T_{\rm e}$), and 
the oxygen abundances estimated using three different ways for the objects in our sample are reported in Table~\ref{tab2}. 
The oxygen abundances of some objects can not be determined through the relations given by SB98 because 
for those objects the estimated abundances or electron densities are outside of the validity ranges considered  
by these authors, i.e., beyond of the range of values for which the relations are defined.
The electron densities, $N_{\rm e}$, for the objects in our sample were computed from the line intensity
ratio  [\ion{S}{ii}]$\lambda 6716/\lambda 6731$ using the {\sc temden} routine of the nebular package
 of {\sc iraf}\footnote{Image Reduction and Analysis Facility, distributed by NOAO, operated by AURA, Inc., under
agreement with the NSF.}.  Most of our objects have values in the range $0.01<n_{\rm e}<0.12$,  
where $n_{\rm e}=N_{\rm e}/(10^{4} \rm cm^{-3})$.
These  values are somewhat higher than the ones derived for \ion{H}{ii} regions, which were found  to be
$n_{\rm e}<0.06$ (see \citealt{krabbe14, copetti00}). 
However, they are sufficiently low so that one does not need to consider the contribution from the collisional de-excitation \citep{rubin89} or the direct 
recombination of the forbidden lines used in the standard abundance determinations.

We found  the electron temperature  $t_{3}$ to be in the range of 6\,000 to 50\,000 K, 
with an average value of  $\sim$20\,000 K.  This  value is somewhat higher than the one found 
by \citet{zhang13} for Sy2s, who used data from SDSS DR7 and found an average value of 14\,000 K.

\begin{figure}
\centering
\includegraphics[angle=-90,width=1\columnwidth]{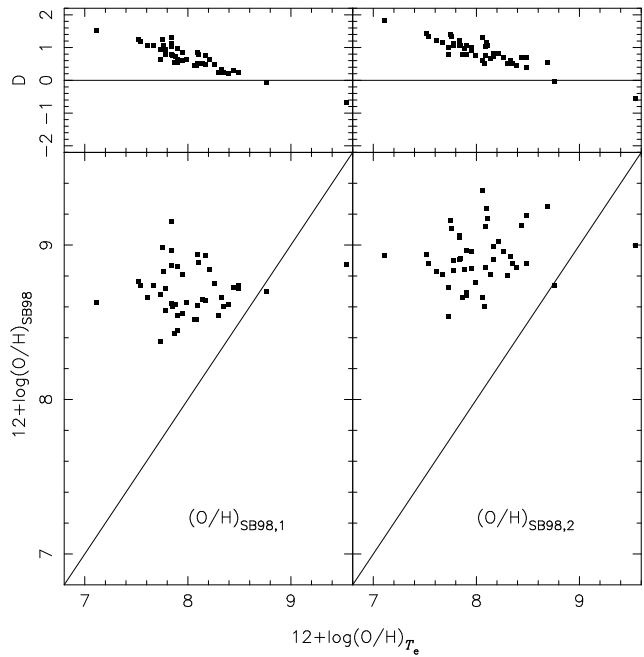}
\caption{The bottom panel shows the comparison between the oxygen abundance obtained through the $T_{\rm e}$ method and 
through the two calibrations of SB98 as indicated in each plot.
The upper panel shows the differences between those O/H estimations.}
\label{f4}
\end{figure} 

Fig.~\ref{f4} shows the comparison between oxygen abundances in NLRs determined through  the two
 methods presented above. 
Inspection of Fig.~\ref{f4} reveals that the $T_{\rm e}$-method provides oxygen abundances lower than 
the ones obtained via the relations of SB98  by up to $\sim$2\,dex,  with an average value of $\sim 0.8$ dex,
which confirms the existence of  the so-called ``temperature 
problem'' in AGNs.  This discrepancy is much larger than the one  found for \ion{H}{ii} regions. For example, 
\citet{kennicutt03}, using spectroscopy data of \ion{H}{ii} regions in spiral galaxies, found that O/H abundances 
computed from relations based on photoionization models of \citet{kewley02} are higher by 
0.2 to 0.5 dex than the ones from the $T_{\rm e}$-method.

The abundance discrepancy in AGNs could be caused by the  presence of a  secondary  heating (ionizing) source in addition to the
radiation from the gas accretion onto the AGN. This secondary source is probably related to the shock.
Indeed,  \citet{zhang13} showed that the  strong [\ion{O}{iii}]$\lambda$4363  flux in AGNs  (and consequently high electron temperature)
suggests the presence of some supplementary energy source(s), which could be due to the presence of shock waves 
(see also \citealt{contini12, prieto05}).  Moreover, the presence of fluctuations of  electron temperature  \citep{peimbert67} in the
gas phase of AGNs could  also contribute for the discrepancies found in Fig.~\ref{f4}.
  
\begin{figure}
\centering
\includegraphics[angle=-90,width=1\columnwidth]{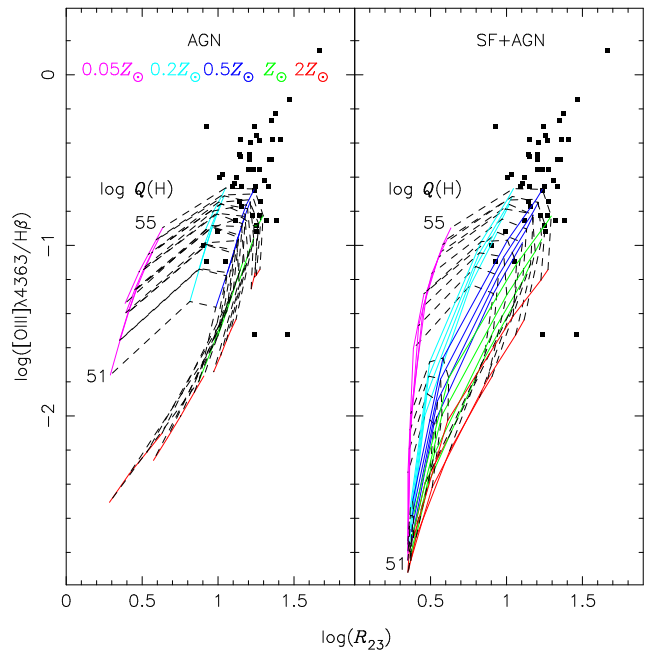}
\caption{The [\ion{O}{iii}]$\lambda$4363/H$\beta$ line ratio as a function of the 
$R_{23}=[I([{\rm O\:II}]\lambda3727)+I([{\rm O\:III}]\lambda4959+\lambda5007)]/I({\rm H\beta})$
 parameter.
The left panel shows the predictions of the  AGN model.  The predictions of 
SF+AGN model are plotted in right panel. The points are the objects of our sample (see Table~\ref{tab1}). 
The typical error (error bars not shown) of the emission-line 
ratios is about 10 per cent.
The solid lines connect the predictions of the models with the same metallicity $Z$,  
while dashed lines connect models with the same ionizing photon number (Q(H)).}
\label{f2ac}
\end{figure}

In order to examine if Sy2 galaxies  have a  secondary  heating   source, we performed a simple test. 
In Fig.~\ref{f2ac}, the intensity
of  [\ion{O}{iii}]$\lambda$4363/H$\beta$ versus  the $R_{23}$ parameter defined as  
$R_{23}=[I([{\rm O\:II}]\lambda3727)+I([{\rm O\:III}]\lambda4959+\lambda5007)]/I({\rm H\beta})$
for the objects in our sample and the ones predicted  by a grid of photoionization models are shown. 

 We consider two models  to reproduce the observational data: 
(i) the  AGN model to describe the NRL spectra, 
and (ii) the SF+AGN model to describe composite (star-forming regions + AGN) spectra. 
 We use the version 08.00 of the Cloudy code \citep{ferland96} to
construct NLR models similar to the ones used by \citet{dors12},  but
 considering the  Table-AGN model \citep{mathews87}  as the ionizing source. 
We  considered a large range of values for the physical parameters: 
the number of ionizing photons,
$51 \lid \:  \log Q({\rm H}) \: \lid \:55$; electron densities, $1.0 \lid  \:  \log(N_{\rm e}) \: \lid 4.0$;
and metallicities,  $\rm 0.05 \: \lid \: (Z/Z_{\odot}) \: \lid \: 2$. 
 The spectrum of the SF+AGN model is the sum of the predictions of the SF model and the AGN model. 
The SF model is constructed using the  Cloudy code \citep{ferland96} assuming the ionizing source  
to be a stellar cluster with an age of 2.5 Myr and $\log Q({\rm H})=52.85$
 whose  spectrum is computed  with the $STARBURST99$ code \citep{leitherer99}.
The  metallicity of the gas phase was considered to be 
 solar  with an electron density $N_{\rm e}=100\, \rm cm^{-3}$. 
  These parameters are similar to the ones  derived  in circumnuclear star-formation regions observed in 
two galaxies (NGC\,1097 and NGC\,6951) containing a Seyfert 2 nucleus by \citet{dors08}.

We can see in  Fig.~\ref{f2ac}  that the ratio [\ion{O}{iii}]$\lambda$4363/H$\beta$ is underpredicted by  the  AGN and SF+AGN
models when compared  the observations.   
 This line ratio depends strongly on the electron temperature, therefore the electron temperatures in AGNs predicted by 
our models are lower than the ones estimated using the $T_{\rm e}$-method. 
Thus, this discrepancy could be attributed to  the presence of gas shock waves propagating
at supersonic velocities through the NLRs, which does  to enhance  the intensity of [\ion{O}{iii}]$\lambda$4363
producing larger electron temperature values and, consequently, low (unrealistic) O/H values when using the $T_{\rm e}$-method
(see \citealt{nagao01} and references therein).

Due to the result above, in our paper the $T_{\rm e}$-method is no longer used to compute the metallicity of AGNs.


\begin{figure}
\centering
\includegraphics[angle=-90,width=1\columnwidth]{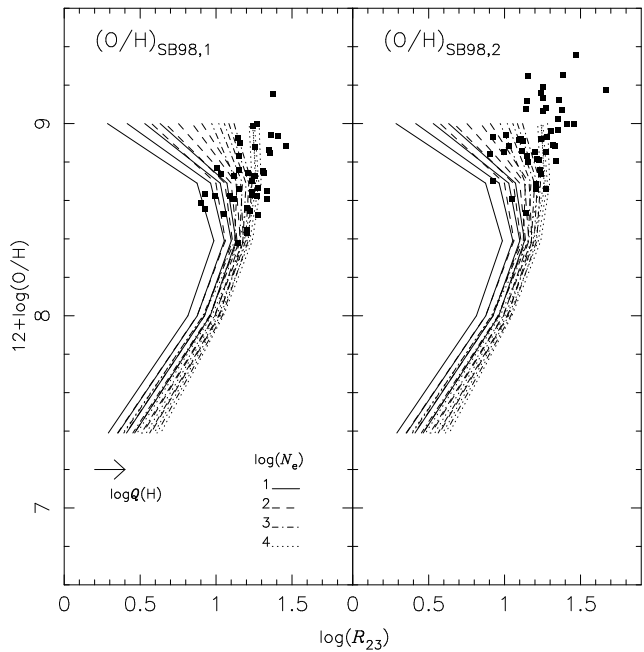}
\caption{12+log(O/H) vs.\ log($R_{23}$). The points represent oxygen abundances in our objects estimated through 
the two relations (see Sect.~\ref{thaisasec}) proposed by SB98 as indicated in each plot.
Lines show our photoionization model results for different values of electron densities, as indicated,
and the logarithm of the ionizing photon numbers ($\log Q$(H)=51-54, with a step of 1 dex).   
The arrow indicates the direction in which the  logarithm of the ionizing photon  numbers  increases 
in  each  model set.
}
\label{f3ac}
\end{figure}

 SB98 pointed out that  an averaged value of $Z$ obtained from both relations  (Eq.\ \ref{equation:sb1} and \ref{equation:sb2}) 
should be used. Instead of doing that we use our models to choose which of the relations by SB98
yields more reliable abundances. For that,  we use the standard O/H -- $R_{23}$ empirical diagram.
 The $R_{23}$ parameter was suggested by \citet{pagel79} as indicator of the oxygen abundance when the
$T_{\rm e}$-method cannot be applied. Although $R_{23}$ has a strong dependence on metallicity, 
it also depends on the ionization degree of the nebulae, which should be taken into account \citep{pilyugin01, leonid00}.  
We compare the predictions of our AGN models with the observed O/H -- $R_{23}$ diagram for the objects in our sample, where the oxygen abundances are 
estimated using both relations of SB98 (see Fig.\ \ref{f3ac}).  Since the AGN and SF+AGN models predict similar values of 
$R_{23}$ in the zone of Fig.\ \ref{f2ac} where our data are located, only the former models are shown in  Fig.\ \ref{f3ac}.
 In this figure we can see that a better agreement is obtained when the abundances 
are determined through the first relation of SB98. Hereafter this relation will be used.

\section{Abundance at the centre of a galaxy vs.\ central  intersect abundance}

\subsection{The data}

It is a widely accepted practice to specify the abundance at the center of a galaxy by 
the central intersect abundance obtained from the radial abundance gradient 
\citep[e.g.,][among others]{VilaCostas1992,Zaritsky1994,vanZee1998,pily04,pilyugin2007,Gusev2012}. 
The radial abundance gradients in the disks of nearby late-type galaxies were recently determined 
by \citet{pilyugin2014,pilyugin2015}. The extrapolation of the radial abundance gradient to the zero galactocentric 
distance gives the  central intersect abundance (O/H)$_{0}$ in  galaxies. 
On the other hand, \citet{ho97} obtained emission-line spectra of the central regions of many 
nearby galaxies. This provides a possibility to estimate the oxygen abundances at 
the centers of those galaxies. 
  The comparison between central oxygen abundance estimated from the spectrum of the center of the galaxy 
with that obtained  from the radial abundance gradient can tell us something about the chemical evolution of 
the central parts of galaxies.  

We compare the central O/H abundances estimated from the spectra of \citet{ho97}
 and the central intersect abundances obtained by \citet{pilyugin2014,pilyugin2015}.  
We selected from the sample of  \citet{ho97} only the galaxies with central spectra classified by \citet{ho97} as  
H\,{\sc ii}-like regions or Seyferts and which are also in the \citet{pilyugin2014,pilyugin2015} list. 
Our  selected sample contains 45  objects (12 AGNs and 33 star-forming regions)   whose 
emission-line intensities and derived oxygen abundances (see below) are listed
in Table~\ref{tab3}.
Fig.~\ref{figure:bptho}  shows the positions of the 45 sample objects in a BPT diagnostic diagram \citep{baldwin81}, 
together with a large sample of emission-line SDSS galaxies studied by \citet{Thuan2010}, which are plotted with cyan (grey) symbols. In this figure we also plotted 
the boundary lines obtained by \citet{Kauffmann2003} and \citet{Kewley2001} that separate the H\,{\sc ii}-like objects and AGNs in the diagram.

\begin{table*}
\caption{De-reddened nuclear emission-line fluxes (normalised to the flux of H$\beta$=1.00) taken from  \citet{ho97} and 
oxygen abundances derived by us. The oxygen abundances 12+log(O/H) for the NLRs of AGNs and central star-forming regions 
are estimated through the SB98 calibration (Eq.~\ref{equation:sb1}) and through the $C_{NS}$ method, respectively.
The values of 12+log(O/H)$_{0}$ are the central intersect abundances in the galaxies obtained from the radial abundance gradients.
}
\label{tab3}
\begin{tabular}{lccccc}	 
\noalign{\smallskip} 
\hline 
 Object            &   [\ion{O}{iii}]$\lambda$5007   & [\ion{N}{ii}]$\lambda$6584 & [\ion{S}{ii}]$\lambda$6717+$\lambda$6731 &   12+log(O/H)       &  12+log(O/H)$_{0}$\\
\hline
      \multicolumn{5}{c}{NLRs of AGNs} \\ 
\hline
  NGC 1058  &    3.64    &	     3.50    &	    1.95    &	 8.58  &     8.62  \\ 
  NGC 1068  &   12.10   &	    2.17    &      0.68    &	 8.54  &     8.64  \\ 
  NGC 2336  &    2.84    &	    5.12     &	    2.99    &	 8.74  &     8.80  \\ 
  NGC 3031  &    4.07    &	    6.38     &	    3.89    &	 8.78  &     8.58  \\ 
  NGC 3227  &    5.92    &	    3.82     &	    1.95    &	 8.62  &     8.64  \\ 
  NGC 3486  &    4.50    &	    3.01     &	    2.64    &	 8.56  &     8.60  \\ 
  NGC 4258  &   10.20   &	    2.29    &	   2.61    &	 8.51  &     8.54  \\ 
  NGC 4395  &   6.23     &           0.93    &	    2.04    &	 8.34  &     8.19  \\ 
  NGC 4501  &   5.15     &	     6.02    &	    2.66    &	 8.80  &     8.92  \\ 
  NGC 4725  &   6.68     &	     3.27    &	    1.82    &	 8.59  &     8.83  \\ 
  NGC 5033  &   4.59     &	     6.76    &	    2.99    &	 8.85  &     8.64  \\ 
  NGC 5194  &   8.09     &	     8.26    &	    2.28    &	 9.01  &    8.88  \\ 
\hline
      \multicolumn{5}{c}{Nuclear star-forming regions} \\ 
\hline
  NGC  598  &      1.57      &         0.48             &        0.61    &   8.35  &	  8.48  \\ 
  NGC  783  &      0.19      &         1.03             &        0.55    &   8.65  &	  8.68  \\ 
  NGC  925  &      0.81      &         0.63             &        0.96    &   8.41  &	  8.48  \\ 
  NGC 1156  &     4.51      &         0.28             &        0.46    &   8.24  &	 8.16  \\ 
  NGC 2403  &     1.91      &         0.79             &        1.03    &   8.36  &	 8.48  \\ 
  NGC 2537  &     1.84      &         0.43             &        0.48    &   8.35  &	 8.35  \\ 
  NGC 2903  &     0.08      &         0.97             &        0.52    &  8.71   &	8.82  \\ 
  NGC 2997  &     0.31      &         1.00             &        0.77    &   8.56  &	  8.80  \\ 
  NGC 3184  &     0.12      &         0.94             &        0.55    &   8.67  &	  8.66  \\ 
  NGC 3198  &     0.23      &         1.19             &        0.84    &   8.59  &	  8.60  \\ 
  NGC 3310  &     0.91      &         1.88             &        0.71    &   8.55  &	  8.37  \\ 
  NGC 3319  &     0.99      &         0.46             &        1.71    &   8.26  &	  8.50  \\ 
  NGC 3344  &     0.83      &         1.20             &        1.15    &   8.49  &	  8.72  \\ 
  NGC 3351  &     0.25      &         1.31             &        0.66    &   8.62  &	  8.82  \\ 
  NGC 3359  &     0.54      &         0.88             &        1.55    &   8.43  &	  8.40  \\ 
  NGC 3631  &     0.29      &         1.23             &        0.68    &   8.60  &	  8.71  \\ 
  NGC 3738  &     2.99      &         0.26             &        0.81    &   8.13  &	  8.10  \\ 
  NGC 3893  &     0.21      &         1.05             &        0.61    &   8.62  &	  8.73  \\ 
  NGC 4088  &     0.19      &         0.91             &        0.48    &   8.65  &	  8.71  \\ 
  NGC 4214  &     3.66      &         0.19             &        0.33    &   8.20  &	  8.20  \\ 
  NGC 4254  &     0.87      &         1.37             &        0.59    &   8.54  &     8.77  \\ 
  NGC 4303  &     1.32      &         2.23             &        1.12    &   8.53  &     8.78  \\ 
  NGC 4449  &     2.36      &         0.40             &        0.65    &   8.26  &     8.26  \\ 
  NGC 4490  &     2.34      &         0.71             &        1.89    &   8.22  &     8.29  \\ 
  NGC 4535  &     0.13      &         1.17	       &        0.61    &   8.66  &	  8.71  \\ 
  NGC 4559  &     0.32      &         1.20	       &        1.10    &   8.53  &	  8.53  \\ 
  NGC 4631  &     1.51      &         0.69             &        0.63    &   8.42  &	 8.39  \\ 
  NGC 4654  &     0.14      &         0.77             &        0.60    &   8.63  &	 8.66  \\ 
  NGC 4656  &     3.96      &         0.14             &        0.50    &   8.04  &	8.06  \\ 
  NGC 5248  &     0.26      &         1.48             &        0.64    &   8.61  &	8.64  \\ 
  NGC 5457  &     0.23      &         1.08             &        0.70    &   8.61  &	8.71  \\ 
  NGC 5474  &     1.75      &         0.37             &        0.72    &   8.26  &	8.19  \\ 
  NGC 6946  &     0.37      &         1.81             &        0.81    &   8.55  &	8.72  \\ 
\hline 
\end{tabular}
\end{table*}

\citet{pilyugin2014,pilyugin2015} derived oxygen abundances from the published emission-line intensities of disk \ion{H}{ii} regions 
trough the  $C$ method \citep[see][]{pily12}. 
This is an empirical method based on the comparison between strong emission-line intensities in the spectrum of a target \ion{H}{ii} region 
and those in a set of  reference \ion{H}{ii} regions with known abundances. The strong emission-lines considered in this method 
are [\ion{O}{ii}]$\lambda$3727, H$\beta$, [\ion{O}{iii}]$\lambda$5007, [\ion{N}{ii}]$\lambda$6584  and [\ion{S}{ii}]$\lambda$6716, $\lambda$6731.
The optical  data of \citet{ho97} consist of long-slit spectroscopy of  the nuclear  region ($r \la 200$ pc) of a large sample of nearby 
galaxies covering the 4200-5200\AA\ and  6200-6900\AA\ spectral ranges with spectral resolutions of about 4 and 2.5 \AA~pixel$^{-1}$, respectively.
We use again the first relation of SB98 to estimate the central abundances of the NLRs of the active galaxies [(O/H)$_{\rm SB98,1}$]. 
In the case of galaxies with central H\,{\sc ii}-like regions we use the $C_{NS}$ method \citep{pily12,pily13} to estimate the oxygen 
abundances [(O/H)$_{C_{NS}}$]. The $C_{NS}$ method is a variant of the $C$ method applicable when the [\ion{O}{ii}]$\lambda$3727 emission-lines are not available.

\begin{figure}
\resizebox{1.00\hsize}{!}{\includegraphics[angle=000]{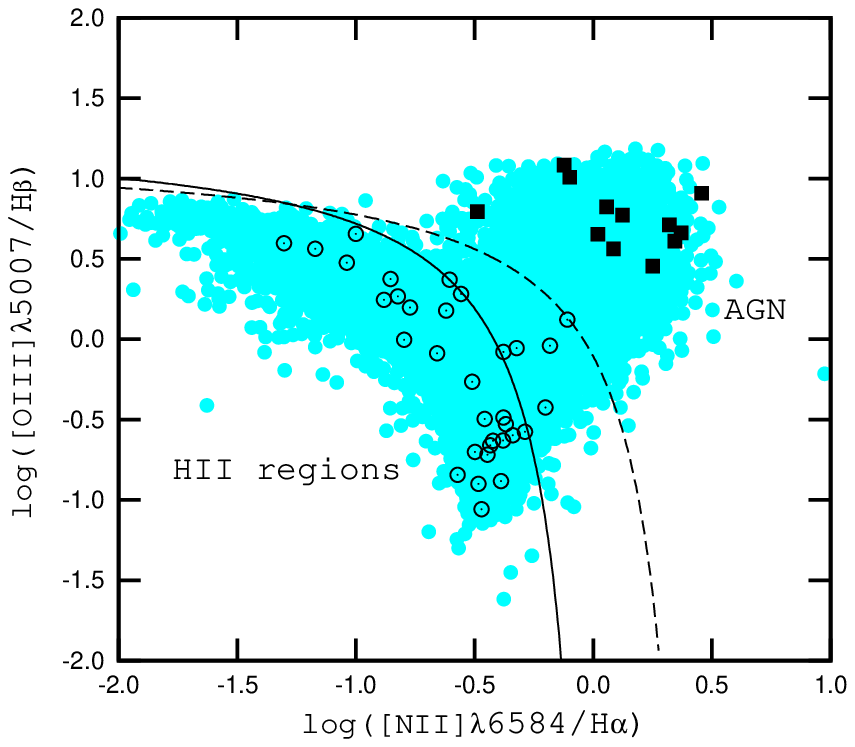}}
\caption{BPT [O\,{\sc iii}]$\lambda$5007/H$\beta$ vs.\ [N\,{\sc ii}]$\lambda$6584/H$\alpha$ diagnostic diagram. 
Filled squares and open circles stand for the AGNs and star-forming central regions selected from the work 
of \citet{ho97}. Solid and dashed curves mark the boundary between AGNs and H\,{\sc ii} regions 
defined by \citet{Kauffmann2003} and \citet{Kewley2001}, respectively.
The filled circles show a large sample of emission-line SDSS galaxies studied by \citet{Thuan2010}. 
}
\label{figure:bptho}
\end{figure}

\begin{figure}
\resizebox{1.00\hsize}{!}{\includegraphics[angle=-90.0]{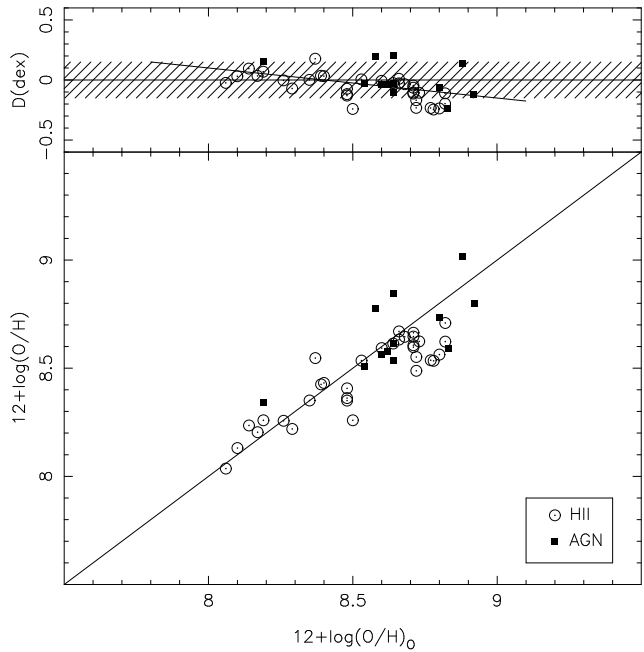}}
\caption{Comparison between central intersect oxygen abundances derived from the radial abundance gradients (O/H)$_{0}$ 
and the central abundances determined from the spectra of the central regions from \citet{ho97} through the SB98 calibration 
for the AGNs (filled squares) and through the $C_{NS}$ method for the H\,{\sc ii}-like regions (open circles). 
The diagonal line in the bottom panel is  the one-to-one relation. 
The upper panel shows the differences between the spectral and central intersect abundances.
The hatched area indicates a band  of $0.15$ dex adoped for the oxygen abundance uncertainty (see Section~\ref{sdata}). 
 Solid line shows the linear regression [$\rm D=(-0.25\pm0.06)\times 12+\log(O/H)_{0} + 2.10 (\pm0.59)$]
considering  all data.}
\label{figure:oho-ohho}
\end{figure}

\subsection{Abundance at the centre  vs. central  intersect abundance}

 In the  bottom panel of Figure~\ref{figure:oho-ohho} we show  the comparison between the central intersect abundances in the 
selected galaxies (O/H)$_{0}$ obtained from the radial abundance gradients and central abundances (O/H)$_{\rm SB98,1}$ and (O/H)$_{C_{NS}}$ 
determined from the spectral measurements of the central regions of \citet{ho97}. 
This figure shows that the (O/H)$_{\rm SB98,1}$ abundances in the NLRs are close to 
or slightly lower than the central intersect abundances in the host galaxies. The central abundances in galaxies with 
central H\,{\sc ii}-like regions show a similar behaviour. 
Thus the metallicity in the NLRs obtained through the relation of SB98 is close to 
the central metallicity of the host galaxy estimated from the radial abundance gradient. 
The mean difference (D) between the direct and intersect central oxygen abundances (see upper panel of Figure~\ref{figure:oho-ohho}) 
for  H\,{\sc ii}-like objects is $-0.11 \pm0.09 $ at high metallicities 
(12+log(O/H)$>$8.5)  and $0.00 \pm0.08 $ at low metallicities (12+log(O/H)$<$8.5). 
The mean difference for AGNs is $-0.01 \pm0.13$ at high metallicities. 
The differences for the high metallicity AGNs seem to
follow a linear regression with a slope of $-0.32(\pm0.33) \: \rm dex^{-1}$. The differences for H\,{\sc ii}-like objects follow a regression 
with a slope of $-0.45(\pm0.23) \: \rm dex^{-1}$ at high metallicities, and
with a slope of $-0.23(\pm0.15) \: \rm dex^{-1}$ at low metallicities.
 We also performed a linear regression considering all points in Fig.~\ref{figure:oho-ohho} due to the small sample we are using. It
yields a slope of $-0.25(\pm0.06) \: \rm dex^{-1}$.

 The differences are within the uncertainty in the abundance estimations (i.e., compatible with zero). 
However, the existence of a trend in the differences suggests that the direct central abundance in some high metallicity galaxies 
can be lower than the central intersect abundance. 
It should be noted that \citet{Sanchez2014} have also found observational evidence of lower central oxygen abundances 
than that predicted by the central abundance extrapolation of the gradients in a number of spiral galaxies.  
A possible explanation of the trend of the differences with metallicity  
could be the accretion of metal-poor gas onto the centers of galaxies. 
In the case of the low-metallicity galaxies, where the metallicity of the accretion material is similar (or not so different)
  to that of the gas in the central region, the infalling gas should not significantly change the local metallicity. 
In contrast, in the case of high-metallicity galaxies this low-metallicity infalling gas would dilute the 
  heavy element content of the gas in the central region. Therefore this effect would be more relevant in 
high-metallicity galaxies than in low-metallicity ones.

Observational evidence of the presence of gas flows from the outer parts (with low-metal content material)  to
the centre of  the galactic disc (high metal content) has been found for isolated barred galaxies 
\citep{martin94, Zaritsky1994} and interacting ones \citep{rosa14, ellison11,  ellison10, kewley10}. 
Likewise, there are several kinematical studies based on optical and infrared 
 integral field spectroscopy that support the scenario where gas is infalling toward the central region of 
AGNs \citep[e.g.][]{muller14, muller11, riffel13b, riffel08, thaisa07, fathi06}. Moreover, \citet{rupke10} 
 performed numerical simulations of galaxy mergers and found that the central underabundances observed in this 
kind of systems could be accounted for by a radial infall of low-metallicity material coming from the 
 outskirts of both galaxies involved.

\subsection{Is there an extraordinary chemical enrichment of the NLRs?}

It was found by \citet{pilyugin2006,pilyugin2007} that there is an upper limit to the oxygen 
abundances in galaxies, i.e., there is a maximum attainable 
oxygen abundance. These authors found  that this maximum value of oxygen abundance in 
galaxies is about twice the solar abundance \cite[adopting the solar oxygen abundance to be  
$\rm 12+\log(O/H)_{\odot}=8.69$;][]{allende01}. 

The abundances in the NLRs of AGNs taken from the list of  \citet{ho97} are in the range  
$\rm 8.5 \: < \: 12+log(O/H) \: <9.0$ or $ 0.6 \: < \:  Z/Z_{\odot} \: < 2$, which does not exceed the maximum attainable 
abundance for galaxies. This suggests that there is no an extraordinary chemical enrichment of the NLRs. 
Similar AGN abundances were obtained by \citet{matsuoka09} from the analysis of the 
C\,{\sc iv}$\lambda$1549/He\,{\sc ii}$\lambda$1640 -- C\,{\sc iii}]$\lambda$1909/C\,{\sc iv}$\lambda$1549 
diagram calibrated in terms of gas abundance through photoionization models using the  {\sc Cloudy} code \citep{ferland96}.
\citet{dors14} also obtained similar abundances using a new index 
(C$_{43}$ = log[(C\,{\sc iv}$\lambda$1549 + C\,{\sc iii}]$\lambda$1909)/He\,{\sc ii}$\lambda$1640])
defined by them as a metallicity indicator for AGNs.

However, it should  be noted that we found metallicities larger than twice solar up to $Z/Z_{\odot}\sim 6.5$ [$12+\log(\rm O/H)_{SB98,1} \sim 9.5$  
(see Fig.\ \ref{f4}]) for five  AGNs from Table \ref{tab1}. Similarly extra high metallicities for some NLRs were obtained by SB98. 
Further study of the galaxies with possible extra high metallicity AGNs should be carried out (in particular, radial abundance 
distributions and central intersect abundances should be obtained in those galaxies) in order to be able to draw solid conclusions 
about the upper limit to the oxygen abundances in NLRs.

\section{Summary and conclusions}

 We compiled from the literature a sample of spectra of narrow-line regions of active galactic nuclei  
with available optical emission lines: 
[\ion{O}{ii}]$\lambda$3727, 
[\ion{O}{iii}]$\lambda$4363, H$\beta$, [\ion{O}{iii}]$\lambda$5007, H$\alpha$, [\ion{N}{ii}]$\lambda$6584,
[\ion{S}{ii}]$\lambda$6717, and [\ion{S}{ii}]$\lambda$6731.
We estimated the oxygen abundances in those NLRs through the classic $T_e$ method and through the strong-line 
method (the calibration of \citealt{thaisa98}). 
We found that the abundances determined through the $T_e$ method are lower by up to $\sim$2 dex than the abundances 
estimated through the calibration of SB98, i.e., we confirmed the existence of the so-called ``temperature problem'' in AGNs. 

 We also considered a second sample of galaxies for which the emission-line spectra of the central regions 
were measured by \citet{ho97} and the central intersect abundances were found by \citet{pilyugin2014,pilyugin2015}. 
The direct central abundances of the AGNs and  Star-forming regions in those galaxies were 
estimated  through the calibration of \citet{thaisa98} and through the $C_{NS}$ 
method  \citep{pily12,pily13}, respectively.
We found that the abundances of the NLRs and H{\sc ii}-like objects estimated from the direct spectral measurements 
are close to or slightly lower than the central intersect abundances obtained from the radial abundance gradients.
This may suggest that the infall of the low-metallicity gas onto the centers of the galaxies can take place in some galaxies 
where the central abundance estimated from the direct spectral measurements is lower than  
the central intersect abundance. 

The abundances in the NLRs of the AGNs in our samples do not suggest that there is an extraordinary chemical enrichment 
of the narrow-line regions.
There are only a few AGNs with oxygen abundances higher than the maximum attainable abundance for galaxies 
\citep[$\sim$2 times the solar value;][]{pilyugin2006,pilyugin2007}. 
Additional investigations of the galaxies with possible extra high metallicity AGNs are necessary to draw reliable
conclusions on the upper limits of the oxygen abundances in NLRs.

\section*{Acknowledgments}
We are grateful to the anonymous referee for his/her useful comments and
suggestions that helped us to substantially clarify and improve the
manuscript.
O.L.D.\ and A.C.K.\ are grateful to the FAPESP for support under grant 2009/14787-7 and 2010/01490-3, respectively.  \\
E.K.G.\ and L.S.P.\ acknowledge support within the framework of Sonderforschungsbereich (SFB 881) on ``The Milky Way System''
(especially subproject A5), which is funded by the German Research Foundation (DFG). \\
O.L.D.\ thanks the hospitality of the Astronomisches Rechen-Institut at Heidelberg University
and the Universidad Nacional de La Plata where part of this work was done. \\
L.S.P.\ thanks the hospitality of the Astronomisches Rechen-Institut 
Heidelberg University where part of this investigation was carried out.  \\
This work was partly funded by the subsidy allocated to Kazan Federal 
University for the state assignment in the sphere of scientific 
activities (L.S.P.).  \\ 
This research has made use of the NASA/IPAC Extragalactic Database (NED) which is operated by the Jet Propulsion Laboratory, 
California Institute of Technology, under contract with the National Aeronautics and Space Administration. \\

\label{lastpage}

\end{document}